\documentclass[12pt,letterpaper]{article} 

\usepackage{osajnl}
\usepackage{overcite,hyperref}

\newcommand{\ud}{\mathrm{d}}
\newcommand{\eqn}[1]{(\ref{#1})}
\newcommand{\fig}[1]{Fig. \ref{#1}}

\begin{document}

\title{Asymptotic Probability Density Function of Nonlinear Phase Noise}

\author{Keang-Po Ho}

\affiliation{StrataLight Communications, Campbell, CA 95008}
\email{kpho@stratalight.com}

\begin{abstract}
The asymptotic probability density function of nonlinear phase noise, often called the Gordon-Mollenauer effect, is derived analytically when the number of fiber spans is very large.
The nonlinear phase noise is the summation of infinitely many independently distributed noncentral chi-square random variables with two degrees of freedom. 
The mean and standard deviation of those random variables are both proportional to the square of the reciprocal of all odd natural numbers. 
The nonlinear phase noise can also be accurately modeled as the summation of a noncentral chi-square random variable with two degrees of freedom and a Gaussian random variable.
\end{abstract}


\maketitle

When optical amplifiers are used to compensate for fiber loss, the interaction of amplifier noise and the Kerr effect causes phase noise, often called the Gordon-Mollenauer effect  or nonlinear phase noise \cite{gordon}.
Nonlinear phase noise degrades both phase-shifted keying (PSK) and differential phase-shift keying (DPSK) systems \cite{ryu, kim} that have renewed attention recently  \cite{gnauck, griffin, zhu}.
Usually, the performance of the system is estimated based on the variance of the nonlinear phase noise \cite{gordon}. 
However, the nonlinear phase noise is not Gaussian noise \cite{kim} and the variance is not sufficient to characterize the system.
The probability density function (p.d.f.) is required to better understand the system and evaluates the system performance. 
This letter provides an analytical expression of the asymptotic p.d.f. for the nonlinear phase noise when the amplifier noise is modeled as a distributed process for a large number of fiber spans.
The characteristic functions are first derived analytically as a simple expression and the p.d.f is the inverse Fourier transform of the corresponding characteristic function.
The asymptotic p.d.f. can be accurately applied to system having more than 32 spans.

For an $N$-span fiber system, the overall nonlinear phase noise is \cite{gordon}

\begin{equation}
\phi_{\mathrm{LN}}  = \gamma L_{\mathrm{eff}} \left\{|A + n_1|^2 + |A + n_1 + n_2|^2 + \cdots + |A + n_1 + \cdots + n_N|^2 \right\},
\label{phinl}
\end{equation}

\noindent where $A$ is a real number representing the amplitude of the transmitted signal, $n_k, k = 1, \dots, N$, are independent identically distributed (i.i.d.) complex zero-mean circular Gaussian random variables as the optical amplifier noise introduced into the system at the $k$th fiber span, $\gamma L_{\mathrm{eff}}$ is the product of fiber nonlinear coefficient and effective fiber length per span. 

With large number of fiber spans, the summation of \eqn{phinl} can be replaced by integration as

\begin{equation}
\phi_{\mathrm{LN}} = \kappa \int_{0}^{L} |A + S(z)|^2 \ud z,
\label{phiint}
\end{equation} 

\noindent where $S(z)$ is a zero-mean complex value Wiener process or Brownian motion of $E\{S(z_1)S^{*}(z_2)\} = \sigma^2_{s} \min(z_1, z_2)$ and $\kappa = N \gamma  L_{\mathrm{eff}}/L$ is the average nonlinear  coefficient per unit length. 
The variance of $\sigma_s^2 = N \sigma^2_{\mathrm{ASE}}/L$ is the noise variance per unit length where $E\{|n^2_k|\} = \sigma^2_{\mathrm{ASE}}, k = 1, \ldots, N $ is noise variance per amplifier.

The p.d.f. is derived for the following normalized nonlinear phase noise

\begin{equation}
\phi = \int_{0}^{1} |\rho + b(t)|^2 \ud t,
\label{phinor}
\end{equation}

\noindent where $b(t)$ is a complex Wiener process with an autocorrelation function of 

\begin{equation}
R_{b}(t, s) = E\{b(s) b^{*}(t)\} = \min(t, s).
\label{cor}
\end{equation}

\noindent Comparing the integrations of \eqn{phiint} and \eqn{phinor}, the normalized phase noise of \eqn{phinor} is scaled by $\phi =  L \sigma_s^2 \phi_{\mathrm{LN}}/\kappa$, $t = z/L$ is the normalized distance, $b(t) = S(tL)/\sigma_s/\sqrt{L}$ is the normalized amplifier noise, and $\rho = A/\sigma_s/\sqrt{L}$ is the normalized amplitude. 
The optical signal-to-noise ratio (SNR) is $\rho^2 = A^2/(L\sigma_s^2) = A^2/(N\sigma^2_{\mathrm{ASE}})$.

The Wiener process of $b(t)$ can be expanded using the standard Karhunen-Lo\'{e}ve expansion of \cite[\S6-4]{root}

\begin{equation}
b(t) = \sum_{k = 1}^{\infty} \sigma_k x_k \psi_k(t),
\label{sum}
\end{equation} 

\noindent where $x_k$ are i.i.d. complex circular Gaussian random variable with zero mean and unity variance, $\sigma_k^2, \psi_k(t),  0 \leq t \leq 1$ are the eigenvalues and eigenfunctions, respectively, of the following integral equation,

\begin{equation}
\psi_k(t) = \sigma_k^2 \int_{0}^{1} R_{b}(t, s) \psi_k(s) \ud s,
\label{int_eq}
\end{equation}

\noindent with boundary condition of $\psi_k(0) = 0$. 
The eigenfunctions of $\psi_k(t)$  are orthonormal 

\begin{equation}
\int_0^1 \psi_k(t) \psi_l(t) \ud t = \left\{
	\begin{array}{ll}
		1 & k = l \\
		0 & k \neq l
	\end{array} \right. .
\end{equation}

Substitute the correlation function of \eqn{cor} into the integral equation of \eqn{int_eq}, we get

\begin{equation}
\psi_k(t) = \sigma_k^2 \int_{0}^{t} s \psi_k(s) \ud s 	+ \sigma_k^2 t \int_{t}^{1} \psi_k(s) \ud s .
\label{int_eq1}
\end{equation}

\noindent Take the second derivative of both sides of \eqn{int_eq1} with respect to $t$, we get

\begin{equation}
\frac{d^2 \psi_k(t)}{d t^2} = - \sigma_k^2 \psi_k(t)
\end{equation}

\noindent with solution of $\psi(t) = \sqrt{2}\sin(t/\sigma_k)$. 
Substitute into \eqn{int_eq} or \eqn{int_eq1}, we find that 

\begin{equation}
\sigma_k = \frac{2}{(2k-1) \pi},  
\psi_k(t) = \sqrt{2} \sin \left[ \frac{(2 k - 1) \pi}{2} t \right].
\label{sigmak}
\end{equation}

Previous studies \cite{ho} are equivalent to the Karhunen-Lo\'{e}ve transform of finite number of random variables of \eqn{phinl} based on numerical calculation. 
While the eigenvalues of the covariance matrix corresponds approximately to $\sigma^2_k$ of \eqn{sigmak}, the eigenvectors always require numerical calculations \cite{ho}.
The assumption of a distributed process of \eqn{phiint} can derive both eigenvalues and eigenfunctions of \eqn{sigmak} analytically.

Substitute \eqn{sum} with \eqn{sigmak} into the normalized phase of \eqn{phinor}, because $\int_{0}^{1} \sin(t/\sigma_k) \ud t = \sigma_k$, we get

\begin{equation}
\phi = \rho^2 +  2 \sqrt{2} \sum_{k = 1}^{\infty} \sigma_k^2 \Re(x_k) +  \sum_{k = 1}^{\infty} \sigma_k^2 |x_k|^2.
\label{sum_x}
\end{equation}
  
\noindent where $\Re(\cdot)$ denotes the real part of a complex number. 
Because $\sum_{k = 1}^{\infty} \sigma_k^2 = 1/2$ (see \cite[\S0.234]{tab}), we get

\begin{equation}
\phi = \sum_{k = 1}^{\infty} \sigma_k^2 | \sqrt{2} \rho +  x_k|^2.
\label{sum_x1}
\end{equation}

The random variable $| \sqrt{2} \rho +  x_k|^2$ is a noncentral $\chi$-square random variable with two degrees of freedom with a noncentrality parameter of $2 \rho^2$ and a variance parameter of $1/2$ \cite[p.44]{proakis}. 
The normalized nonlinear phase noise is the summation of infinitely many i.i.d. noncentral $\chi$-square random variables with two degrees of freedom with noncentrality parameters of $ 2 \sigma_k^2 \rho^2$ and variance parameter of $\sigma_k^2/2$.
The mean and standard deviation of the random variables are both proportional to the square of the reciprocal of all odd natural numbers. 

The characteristic function of  $| \sqrt{2} \rho +  x_k|^2$  is \cite[p.44]{proakis}

\begin{equation}
\Psi_{| \sqrt{2} \rho +  x_k|^2}(j \nu) = \frac{1}{1 - j \nu} 
	\exp \left( \frac{ 2 j \nu \rho^2} {1 - j \nu} \right),
\end{equation}

\noindent and with  mean and variance of  $2 \rho^2 + 1$ and $4 \rho^2 + 1$, respectively. The characteristic function of the normalized phase $\phi$ of \eqn{phinor} is

\begin{equation}
\Psi_\phi(j \nu) = \prod_{k=1}^{\infty}
	\frac{1}{1 - j \nu \sigma_k^2} 
	\exp \left( \frac{ 2 j \nu \rho^2 \sigma_k^2} {1 - j \nu \sigma_k^2} \right).
\label{eig_phi}
\end{equation}

\noindent Using the expressions of \cite[\S1.431, \S1.421]{tab}, the characteristic function of \eqn{eig_phi} can be simplified to

\begin{equation}
\Psi_\phi(j \nu) = \sec(\sqrt{j \nu}) \exp \left[\rho^2  \sqrt{j \nu} \tan( \sqrt{j \nu} ) \right].
\label{cfsimp}
\end{equation}

The first eigenvalue of \eqn{sigmak} is much larger than other eigenvalues. 
The normalized phase of \eqn{sum_x} is dominated by the noncentral $\chi$-square random variable corresponding to the first eigenvalue because of  

\begin{equation}
\frac{\sigma_1^2}{\sigma_2^2 + \sigma_3^2 + \cdots }
 	= \frac{(2/\pi)^2 }{1/2 - (2/\pi)^2} = 4.27,
\end{equation}

\noindent and

\begin{equation}
\frac{\sigma_1^4 }{\sigma_2^4 + \sigma_3^4 + \cdots } =
\frac{(2/\pi)^4 }{1/6 - (2/\pi)^4}
= 68.12.
\end{equation}

\noindent The relationship of  $\sum_{k=1}^{\infty} \sigma_k^4 = 1/6$ is based on \cite[\S0.234]{tab}.

Beside the noncentral $\chi$-square random variable corresponding to the largest eigenvalue of $\sigma_1$, the other $\chi$-square random variables of $| \sqrt{2} \rho +  x_k|^2$, $k > 1$, have more or less than same variance. From the central limit theorem \cite[\S5-4]{root}, the summation of many random variables with more or less the same variance approaches a Gaussian random variable. The characteristic function of \eqn{eig_phi} can be accurately approximated by

\begin{eqnarray}
\Psi_\phi(j \nu) & \approx & 
	\frac{1}{1 - 4 j \nu /\pi^2} 
	\exp \left( \frac{ 8 j \nu \rho^2/\pi^2} {1 - 4 j \nu /\pi^2} \right) \nonumber \\
	& & \times \exp \left[  j \nu (2 \rho^2 + 1) 
		\left(\frac{1}{2} - \frac{4}{\pi^2} \right)  
	- \frac{1}{2} \nu^2 (4 \rho^2 + 1) 			
	\left( \frac{1}{6} - \frac{16}{\pi^4} \right) 
		\right] ,
\label{eig_phi1}
\end{eqnarray}

\noindent as a summation of a  noncentral $\chi$-square random variable with two degrees of freedom and a Gaussian random variable. 
While the characteristic function of \eqn{cfsimp} is a simpler expression than that of \eqn{eig_phi1}, the physical meaning of \eqn{cfsimp} is more obvious.

The p.d.f. of the normalized phase noise of \eqn{phinor} can be calculated by taking the inverse Fourier transform of either the exact \eqn{cfsimp} or the approximated \eqn{eig_phi1} characteristic functions. 
\fig{pdfsnr} shows the p.d.f. of the normalized nonlinear phase noise for three different optical SNR of $\rho^2 = 11, 18,$ and $25$, corresponding to about an error probability of $10^{-6}$, $10^{-9}$, and $10^{-12}$, respectively, when amplifier noise is the only impairment. 
\fig{pdfsnr} shows that the p.d.f. using the exact \eqn{cfsimp} or the approximated \eqn{eig_phi1} characteristic function, and the Gaussian approximation with mean and variance of $m_\phi = \rho^2 + 1/2$ and $\sigma_\phi^2 = (4\rho^2 + 1)/6$. 
The exact and approximated p.d.f. overlap and cannot be distinguished with each other. 

\fig{pdferf} shows the cumulative tail probabilities as a function of $Q$-factor. 
The $Q$-factor is defined as $Q = (\phi - m_\phi)/ \sigma_\phi$ and gives an error probability or tail probability of $\frac{1}{2} \mathrm{erfc} (Q/\sqrt{2})$ for Gaussian distribution, where $\mathrm{erfc}(\cdot)$ is the complementary error function.
\fig{pdferf} is plotted for the case of $\rho^2 = 18$.
From \fig{pdferf}, the p.d.f. calculated from the exact \eqn{cfsimp} or approximated \eqn{eig_phi1} characteristic function has no difference. 
The Gaussian approximation underestimates the cumulative tail probability for $Q > 1$ but overestimates the cumulative tail probability for $Q < 1$.

The p.d.f. for finite number of fiber spans was derived base on the orthogonalization of \eqn{phinl} by $N$ i.i.d. random variables \cite{ho}.  
\fig{pdfN} shows a comparison of the p.d.f. for  $N = 4, 8, 16, 32$, and $64$ of fiber spans \cite{ho} with the distributed case of \eqn{cfsimp}. 
Using an optical SNR of $\rho^2 = 18$, \fig{pdfN} is plotted in logarithmic scale to show the difference in the tail.
\fig{pdfN} also provides an inset in linear scale of the same p.d.f. to show the difference around the mean.
The asymptotic p.d.f. of \eqn{cfsimp} with distributed noise has the smallest spread in the tail as compared with those p.d.f.'s with $N$ discrete noise sources. 
The asymptotic p.d.f. is very accurate for $N \geq 32$ fiber spans.  

In summary, this letter derives the asymptotic p.d.f. of nonlinear phase noise when the number of fiber spans is very large. 
Gaussian approximation based solely on the variance cannot use to predict the performance of the system accurately.
The nonlinear phase noise can be modeled accurately as the summation of a noncentral $\chi$-square random variable with two degrees of freedom and a Gaussian random variable.

\newpage

\section*{List of Figure Captions}

\begin{figure}[h]\centerline{\scalebox{0.75}{\includegraphics{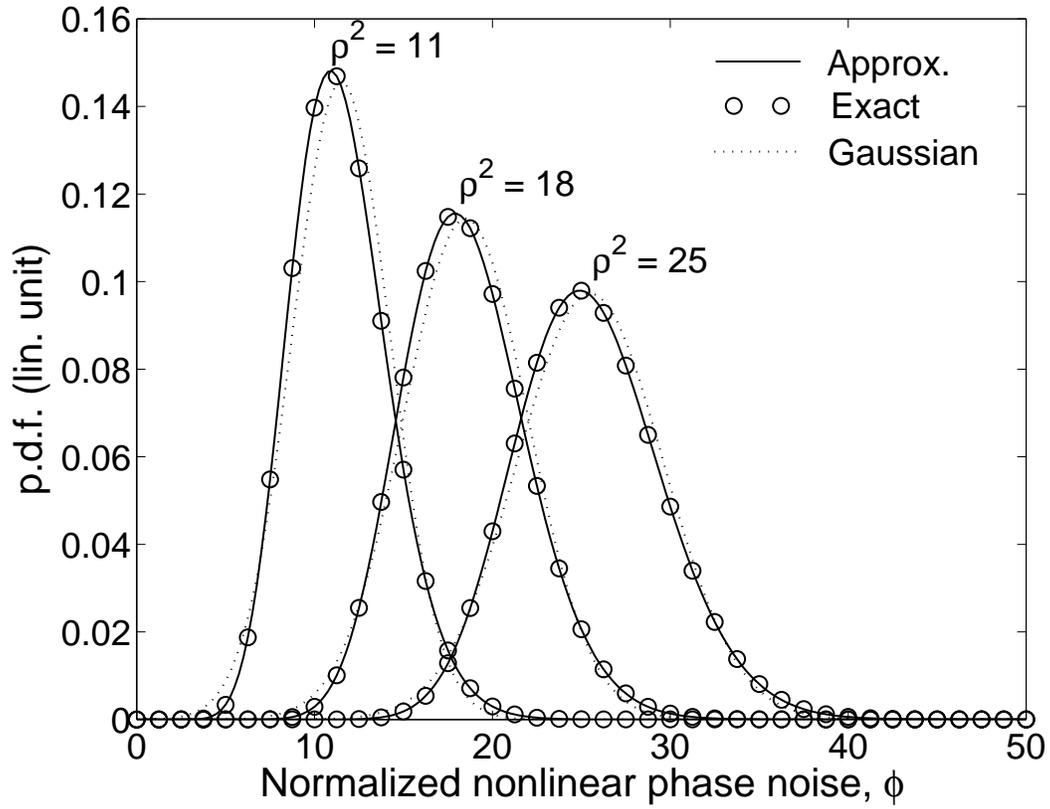}}}
\caption{The p.d.f. of the normalized nonlinear phase noise $\phi$ for optical SNR of $\rho^2 = 11, 18,$ and $25$.}
\label{pdfsnr}
\end{figure}

\fig{pdfsnr}. The p.d.f. of the normalized nonlinear phase noise $\phi$ for optical SNR of $\rho^2 = 11, 18,$ and $25$.

\begin{figure}[h]\centerline{\scalebox{0.75}{\includegraphics{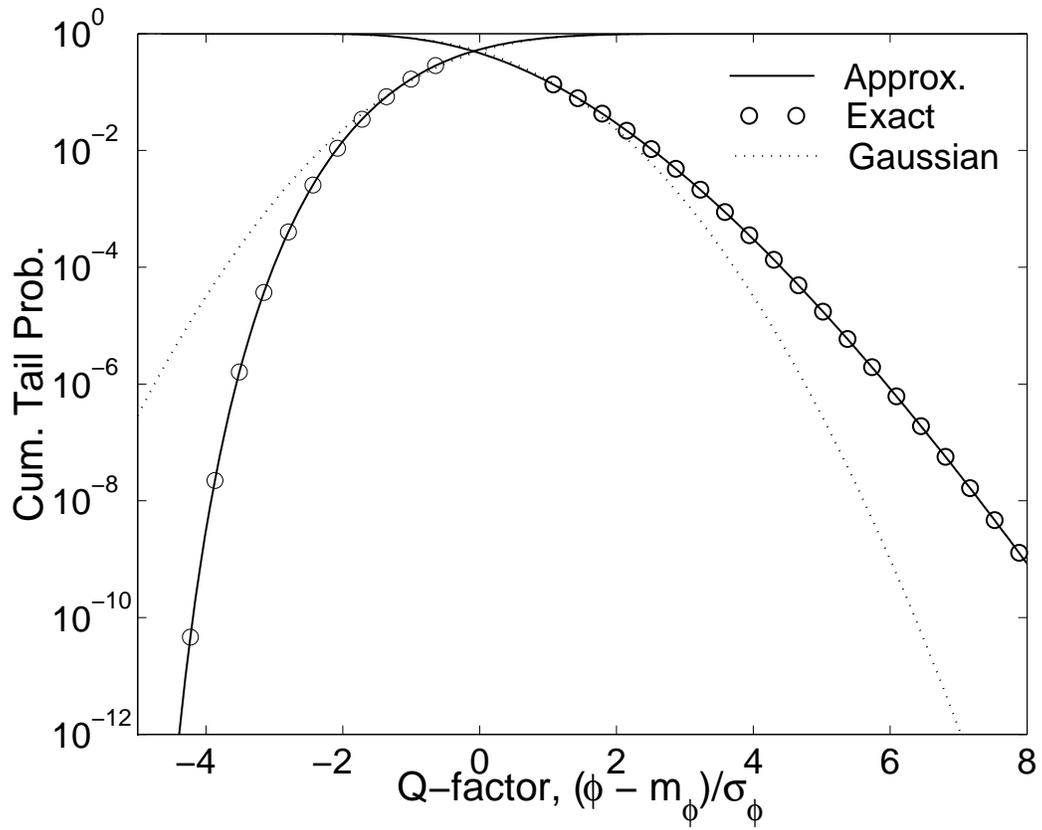}}}
\caption{The cumulative tail probability as a function of $Q$-factor.}
\label{pdferf}
\end{figure}

\noindent \fig{pdferf}. The cumulative tail probability as a function of $Q$-factor.

\begin{figure}[h]\centerline{\scalebox{0.75}{\includegraphics{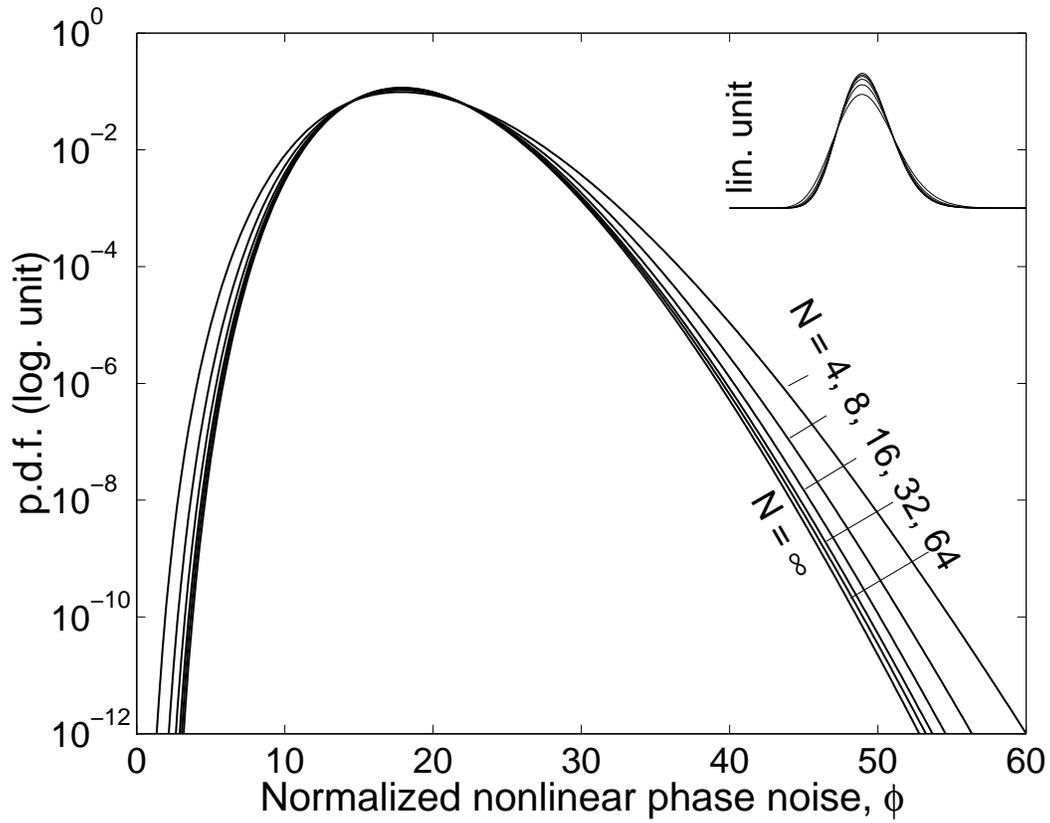}}}
\caption{The asymptotic p.d.f. of $\phi$ as compared with the p.d.f. of $N=4, 8, 16, 32,$ and $64$ fiber spans.
The p.d.f. in linear scale is shown in the inset.}
\label{pdfN}
\end{figure}

\noindent \fig{pdfN}. The asymptotic p.d.f. of $\phi$ as compared with the p.d.f. of $N=4, 8, 16, 32,$ and $64$ fiber spans.
The p.d.f. in linear scale is shown in the inset.

\newpage

\end{document}